# Highly efficient acoustic refractive metasurfaces by harnessing near field coupling

Zhilin Hou,[1]† Xinsheng Fang,[2]† Yong Li,[2]* Badreddine Assouar[3]*

[1]School of Physics and Optoelectronics, South China University of Technology, Guangzhou 510640, China.
[2]Institute of Acoustics, School of Physics Science and Engineering, Tongji University, Shanghai 200092, China.
[3]Institut Jean Lamour, Université de Lorraine, CNRS, F-54000 Nancy, France.

† These authors contributed equally to this work.
* yongli@tongji.edu.cn
* badreddine.assouar@univ-lorraine.fr

## Abstract

Typical acoustic refractive metasurfaces governed by generalized Snell's law require several types of subwavelength subunits to provide an extra phase gradient along the surface. This design strategy, however, has several kinds of drawback. For instance, the inevitable viscous loss brought out by the complex subwavelength subunits, and the negligence of the coupling between adjacent subunits which leads to low-efficiency in wavefront manipulation, especially for large angles. To overcome these limitations, we propose a new type of refractive metasurface composed of only one straight channel and several surface-etched grooves per period. By harnessing the nonlocal coupling between the channel/grooves, and the evanescent modes inside them, superiorly feasible acoustic transmission manipulations can be achieved. Nearly perfect acoustic bending with transmission efficiency up to 95% is demonstrated with theory and experiments for an extremely large angle of 81°. The reported results introduce a novel concept of acoustic metasurfaces and offer a real leap towards the development of high-efficient acoustic devices for wavefront manipulation.

## Introduction

Highly efficient wave manipulation via artificial structures is strongly desired in materials physics and engineering communities. As a thin compact structure, acoustic metasurfaces has attracted significant attention in recent years because of their unique functionalities and capabilities in controlling and transforming the wavefront [*1*]. Many fascinating and exotic functionalities, such as anomalous refraction and reflection [*2-10*], asymmetric transmission [*11-13*], retroreflection [*14, 15*], perfect absorption [*16, 17*], and other abnormal wave phenomena [*18-20*] have been realized by using such kind of inhomogeneous two-dimensional materials.

It can be found however that, the devices to realize those functionalities and physical effects are mostly constructed based on the phase-gradient approach [*21*], which often provide very complicated building subunits. For example, to design a metasurface by which an incident wave can be steered into an anomalous direction, we first should predetermine a continuous surface with suitable lateral phase profile, and then discretize the phase profile into subwavelength unit-cell sizes and implement them by intentionally designed elements. To improve the lateral discretizing resolution and to meet at the same time the requirement of matching both the phase and amplitude profiles of the input and output fields, the subunits have to be designed as multi-folded tubes [*8*] or multi-connected Helmholtz resonators [*5*] with narrow channels and thin walls. Such a complex structure will increase the difficulty in fabrication process and practical implementation. More importantly, because the viscous effect in narrow channels and deformations of the thin walls become larger when their size becomes thinner, we hardly can design a metasurface for wave manipulation in small wavelength range. These important drawbacks will greatly limit their applications in innovative technologies.

Recently, metasurfaces for anomalous refraction were revisited by several groups for both electromagnetic [*22, 23*] and acoustic waves [*2, 3*]. These works are originally raised to improve the efficiency of metasurfaces. It has been pointed out that, because the nonlocal effect caused by the surface evanescent mode has not been taken into account in the design procedure, devices based on phase-gradient approach have intrinsic problem of efficiency. To solve it, the nonlocal effect was suggested to be pre-included into an impedance matrix profile, upon which the meta-atoms can be designed [*22*]. With these efforts, the efficiency of metasurfaces has been improved. However, to realize the impedance matrix profile along a surface, one still need to discretize it into sub-units. This mean narrow channels and thin walls will still need to be used in the subunits structure. Furthermore, because the transmission and reflection coefficients of meta-atoms are required to be bi-anisotropic, the design of meta-atoms for such kind of structure becomes more complicated.

In this paper, we show alternatively that a refractive metasurface can also be designed as simple as a periodic sound-hard planar layer with one straight-walled channel and several surface-etched grooves per period. We name it as single meta-atom metasurface because there is only one channel per period in the structure. In the latter, we do not need a discretizing and matching procedure in the design procedure, which means, structure with very narrow channel and very thin wall can be avoided. As demonstrations, we show that, a structure with one channel and several grooves can perfectly refract a normal incident plane wave into directions with angles as 63°, 72° and 81°. The obtained structures with transmission efficiency over 95% are checked by finite element simulation supplied by Comsol Multiphysics. The physics for those structures is also investigated. We find that the power flow can be strongly redistributed in the lateral direction by the surface evanescent mode. These results suggest that the local power conservation condition in the surface-normal direction, which is considered as the basic requirement in the bi-anisotropic metasurface designing, is in fact unnecessary to be satisfied. The sample with refractive angle as 72° is further evidenced by experiment. The measured data perfectly agreed with the simulation results.

It is necessary to point out that, although perfect metasurfaces with simple structure for electromagnetic waves have been reported recently [24-26], the control of sound would however result in bulky device if we directly translate the same mechanism into acoustics. Currently, efforts have been made on reflective metasurfaces for acoustic waves, and devices with single subunit [27] or with simply structured subunits per period [9, 28] have been suggested, but a refractive one with simple structure is still a challenge.

## Results

**Structure of the metasurface and the Grating theory for diffraction property**

We consider the structure plotted schematically in Fig.1, which is a periodic planar sound-hard layer with thickness $h$ and lateral pitch $a$. In each period, there are one straight-walled channel with width $t_c$ and $L$ rectangular-shaped grooves on both the upper and lower surfaces ($L=2$ in the figure is show). To make the structure as compact as possible in thickness, and for sake of simplicity, in this research we set all the widths of these grooves to be $t$ and all of intervals between to be $w$. The depth of the $l$th ($l=1,\cdots,L$) grooves is denoted by $d_l^{(u,d)}$, where the superscript $u$ $(d)$ means the upper (lower) surface. Notice that these depths should satisfy the condition $d_l^u + d_l^d < h - d_w$ for the considered structure, where $d_w$ is the smallest thickness of the wall between the upper and lower grooves.

As is shown by blue arrows in Fig. 1(A), the purpose of our designing is to refract the normally incident energy into an angle $\theta_t$ with 100% efficiency. It has been proven in Refs.[3] and [22] that this purpose cannot be fulfilled by the phase-gradient metasurfacce because one cannot design such a subunit series, with which the required phase gradient along the metasurfaces and the impedance matching between the incident and the desired scattered waves could be simultaneously satisfied. We argue that the main physics behind this difficult stems mainly from the effect of near field interaction. As we know, as the subunits are isolated with each other, we can indeed manage their transmission and phase delay simultaneously by the structure, but when they are closely gathered in one period in wavelength scale, they will interact strongly with each other. As a result, the physical properties of these subunits depend strongly on their neighbors, or say, their properties are nonlocal. Based on this understanding, we select the general grating theory to solve the problem. The advantage of this theory is that it can include explicitly the high-order evanescent modes, by which the interaction between the subunits can be fully taken into considered in the design procedure.

According to the general grating theory, an incident wave from the negative *y*-direction will be diffracted as 0th, ±1st, ⋯, order diffractive components on both side of the structure. The design procedure can be started by adjusting the relative position, the width, and the depth of the grooves and channel, so that the structure can extinguish all the other propagation components except the one in the desired direction. However, the problem can hardly be solved in a direct way because there are too many parameters in the structure. In this paper, we employ the grating theory combined with an optimization algorithm to solve this problem. In addition, to guarantee a high compactness of the device and to simplify the design procedure, we restrict our research just on the configuration shown in the figure.

To design such a metasurface, we develop a half-analytical method based on the general grating theory, by which the diffractive property of a structure with given channel and grooves can be rigorously calculated. With this method, the desired structure is searched by an optimization algorithm. The detail of the method and the description of the optimization procedure can be found in the supplemented material.

**Performance of the refractive metasurfaces**

As demonstration, metasurfaces with $\theta_t=63°$, 72° and 81° under normal incident plane wave are searched, where $\theta_t$ is the angle between the direction of transmitted wave and surface normal. For these purposes, we first set the period of the structure as $a=\lambda/sin\theta_t$, under which only the 0th and ±1st order diffractive components on both side of the structure are propagating modes, that means what we need to do is to find a suitable geometric substructure, by which the 1st order mode in transmitting end is enhanced and in the same time all the other propagating modes are extinguished. For sake of simplicity, we fix $h=0.4\lambda$, $d_w=0.05\lambda$, $t_c=0.1a$ and $w=0.05a$ for all the structures, and make the optimization only for the depth of grooves. The *L* value is chosen manually as small as possible to get the simplest structure. Under these setting, structures with $L=6$ for $\theta_t=63°$, and $L=5$ for $\theta_t=72°$ and 81° are found. For example, to refract the normally incident plane into the direction with $\theta_t=72°$, the depth of the grooves $d_1^{u(d)} - d_5^{u(d)}$ need to be [0.178(0.123), 0.203(0.017), 0.180(0.000), 0.220(0.000), 0.032(0.288)]$\lambda$, respectively. Structural parameters of the structure for $\theta_t=63°$ and 81° are listed in Tab. S1 in the supplementary material.

To verify the quality and efficiency of the obtained structures, full wave simulation based on the Comsol Multiphysics is performed. Here we evaluate the quality by the value $I_{1y}/I_{ty}$, and evaluate the efficiency by $I_{1y}/I_{iy}$, where $I_{1y}$, $I_{ty}$ and $I_{iy}$ means respectively the intensity of the +1st diffraction, the total transmission and the incident wave. We show in Fig. 2 the real part of the scattering pressure field, where Fig 2(A), (B) and (C) are for the structures with $\theta_t=63°$, 72° and 81°, respectively. For clear eyesight, the field of the normal incident wave is not shown. From the simulation, the efficiency for Fig.2(A)-(C) is obtained as 95.6%, 98.6% and

95.4%, respectively, which are closer to the ones obtained by the numerical calculation. To see the quality of the transmitted wave, we perform a Fast Fourier Transformation of the field along the cut line away from the lower surface 4a. The results show that there are only tiny kinks at $k_x/(k_0 \sin\theta_t)=0$ and -1, which means the power flow is almost completely in the desired direction for all three structures (see the details in the supplemented material). We have also checked the amplitudes of the pressure field in the grooves, and found that the maximum value in all grooves is less than 10 times of the incident one, which means there is no strong resonance in the structure. All these results show that the perfect metasurfaces with only a single meta-atom per period are obtained.

**Power flow redistribution along the surfaces caused by the evanescent mode**

Because there is only one channel per period in the structure, there should be strong evanescent modes along the surfaces, by which the power flow is squeezed into the channel in the input side and then is spread out with suitable phase delay along the surface in the output side. To see this effect, we choose to calculate the local intensity distribution of the field shown in Fig. 2(B), which is for the structure with $\theta_t=72°$. From the field by the finite element simulation, we calculate the local intensity vector $\boldsymbol{I}=(I_x, I_y)$, where $I_i$ ($i=x, y$) can be obtained by the formula $I_i = \frac{1}{2}\operatorname{Re}\left[p \cdot (v_i)^*\right]$ with $p$ as the local pressure value and $(v_i)^*$ as the conjugation of the local velocity component in $i$-direction. Result in one period in $x$-direction and in $a/2$ away from the surfaces in $y$-direction is shown in Fig. 3(A). We can observe that the amplitude of local intensity is shown by the length of the arrow, and the direction of arrow gives the direction of the local intensity. It can be found from this figure that, the directions of the arrows are strongly distorted in the area very close to surface, showing the manner of how the power flow is squeezed into and spread out from the channel. However, as the distance away from the surfaces becomes larger, the distortion becomes smaller and finally gives a regular pattern (about $a/2$ away from the interfaces). Notice the fact that the regular pattern of arrows gives the power flow distribution of the incident or the desired refractive plane waves in far field ($y>a/2$), while the distortion near the interfaces means exactly the lateral energy exchanging caused by the surface evanescent mode.

The figure shows a new manner of how the power flow can pass through the structure, which is quite different from the one in the structures constructed by the discretizing and matching procedure. In the latter, the idea is to discretize and implement the surface with as many as possible meta-atoms (or say, channels) per period, because it is admitted that the power conservation condition should be locally satisfied along the whole surface. Here, we show that the power flow can pass through the metasurface by a completely different way. This means the local power conservation should be an over-requested condition in the previous metasurfaces designing (the phase-gradient and bi-anisotropic approaches), and should be one of the causes for the complexity of the structure.

To show more details about the effect of the evanescent mode, we plot in Fig. 3(B) and (C) respectively the $I_y$ value as the function of $x$ in the upper and lower half-infinite mediums at different $\delta_y$. Here $\delta_y$ means the distance away from the upper or lower interfaces. Results for $\delta_y$=0, 0.05$a$, 0.1$a$, 0.5$a$ and $a$ are shown by different colors in the figures. To guide the eyesight, the position of the channel and grooves is also marked by green and blue rectangles, respectively. It can be seen from the figure that the effect of the evanescent mode is limited in the region around $\delta_y$ <0.5$a$. For all these curves in this region, the $I_y$ value keeps negative in the position where the channel is connected, but keeps dropping from positive to negative in the positions where the grooves are connected. This is understandable because power flow can only pass through the structure from the channel, and the effect of the grooves is only to laterally transfer the power flow along the surface. From these two figures, we can confirm further that the local power conservation condition for is unnecessary.

**Experiment verification**

To verify the numerical result, we select the structure with $\theta_t$ =72° for the experimental demonstration. The schematic representation of the experimental setup is shown in Fig. 4(A). In experiment, we choose the air as working medium and the working frequency as $f$=8200Hz ($\lambda$=42.88mm and $a$=45.08mm). Under this frequency, the width of the channel is 4.5mm, and the width of the grooves and walls are 5.4mm and 2.3mm, respectively. This means that the additional effects of friction and wall deformation caused by the narrow channel and thin wall can be neglected. This frequency is much higher than the one (usually about 3000Hz) used for structures suggested in previous literature. We point out that, because narrow channels and thin walls (compared to the working wavelength) have to be used in the structures suggested in previous works, it is very difficult to push their working frequency into the region as high as 8200Hz.

We show in Figure 4(B) the real part of the pressure field distribution of the structure. The upper panel presents the simulated results and the lower panel presents the corresponding experimental one measured in the areas marked by red boxes in the upper panel. A good agreement between the simulation and experimental results is obtained. Notice that only the field below the metasurface is shown.

**Discussion**

In conclusion, we have developed a half-analytic method to solve the diffraction problem for a periodic sound-hard planar layer with channel and surface-etched grooves. By combining this method with an optimization algorithm, acoustic refractive metasurface which can refract the incident wave into desired directions can be designed. As demonstration, refractive metasurfaces with refractive angles $\theta_t$ =63°, 72° and 81° for normally incident plane wave were designed. The predicted structures were investigated by the finite-element simulation and experiment, which provide a proof-of-concept of the designed transmitting metasurface. In contrast with the traditional refractive metasurface which are usually composed of many subwavelength meta-atoms, the structure designed by our method has only a single meta-atom per period. By this structure, we have found that the local power conservation condition along the surface-normal direction, which is considered as the basic requirement in the previous traditional designs, is in fact an over-requested condition which increases the complexity in metasurface design. Because the near field interaction between subunits exists in almost all of the subwavelength structure, the presented idea and developed method will be helpful for other metasurface designs and functionalities.

## Materials and Methods

### Numerical simulation

The full wave simulations based on finite element analysis are performed using COMSOL Multiphysics Pressure Acoustics module. For Fig.2, plane wave along $-y$ direction is chosen as incident wave. The Perfectly Matched Layers (PML) with thickness $2a$ are added at the top regions (not shown in the figure) to reduce the reflection on the boundaries. Floquet periodic boundary condition is added on the left and right boundaries. For the upper panels of Fig. 4(B), the equal-amplitude beam with finite width in x direction is chosen as incident wave. The plane wave radiation boundary condition on the top, left and right boundaries are used.

### Experimental apparatus

The samples with 20 periods are fabricated using the stereo lithography apparatus (SLA) with photosensitive resin. The molding thickness of each layer during printing is 0.1mm. Organic Glass plates are added on the top and bottom of the samples (height is 15mm) to form a two-dimensional wave guide for measurement. Foams are distributed on sides of the waveguide to absorb sound wave with frequency above 3000Hz. Transducer array includes 22 loudspeakers (1-inch, Hivi B1S) , which can produce continuous sound wave from 6.4kHz to 9.6kHz, is placed 150mm away from the sample as the beam source. The amplitude and phase shift, are measured with Brüel & Kjær Data Acquisition Hardware (LAN-XI, type 3160-A-042) and 1/8-inch microphone controlled by the two-dimensional sweeping field platform with a step of 5mm (about $\lambda/8$) .

**Acknowledgments**

This work was supported by the National Natural Science Foundation of China (Grant Nos. 11274121, 11704284 and 11774265), the Young Elite Scientists Sponsorship by CAST (Grant No. 2018QNRC001), the Shanghai Science and Technology Committee (Grant No. 18JC1410900), the Shanghai Pujiang Program (Grant No. 17PJ1409000), and the Stable Supporting Fund of Acoustic Science and Technology Laboratory.

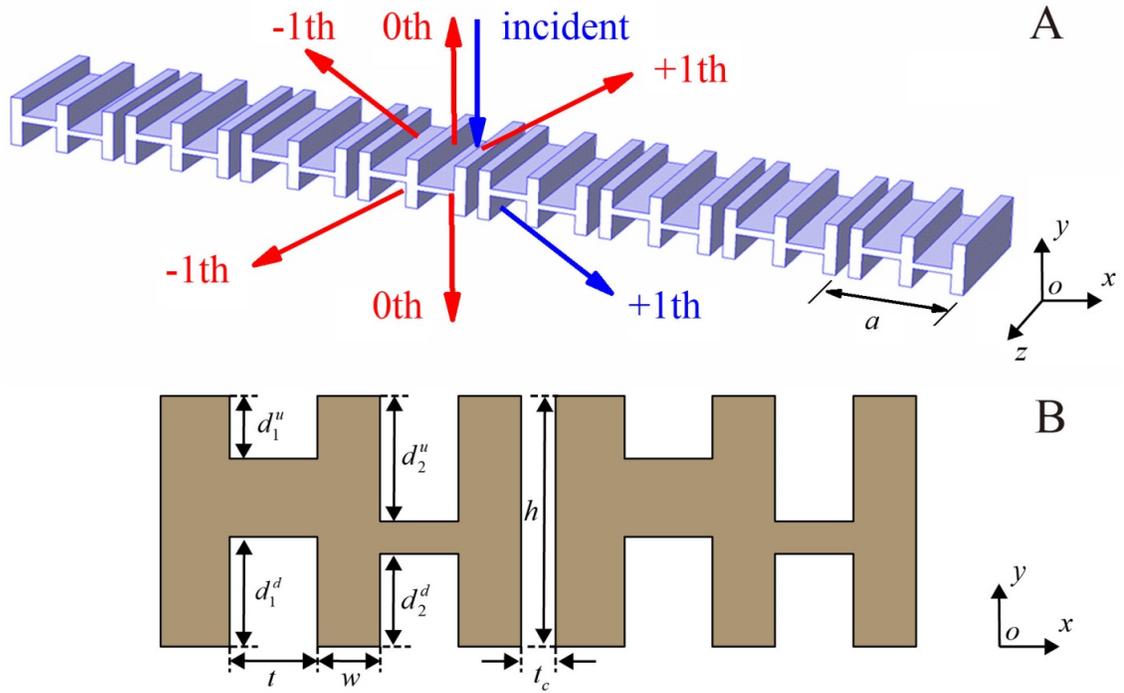

**Fig. 1. Schematic illustration of the investigated structure.** Schematic illustration of the investigated structure. **(A)** three dimensional view and **(B)** detail structure of two periods in xy-cut plane. It is a planar period sound hard surface with one channel and $L$ etched rectangular grooves ($L=2$ is shown in the figure) on both side of layer. The period along $x$-direction is $a$. The total thickness of the layer is $h$, the width of the channel is $t_c$; the width of all of the grooves are set to be the same and denoted as $t$, the depth of the grooves on the upper (lower) surface are $d_l^{u(d)}$, ($l=1,2,…, L$), and the distance between nearest grooves is set to be equal and denoted as $w$. Because of the periodicity of the structure, an incident wave from the negative $y$-direction will be diffracted as 0th, 1st ,⋯, order diffractive components.

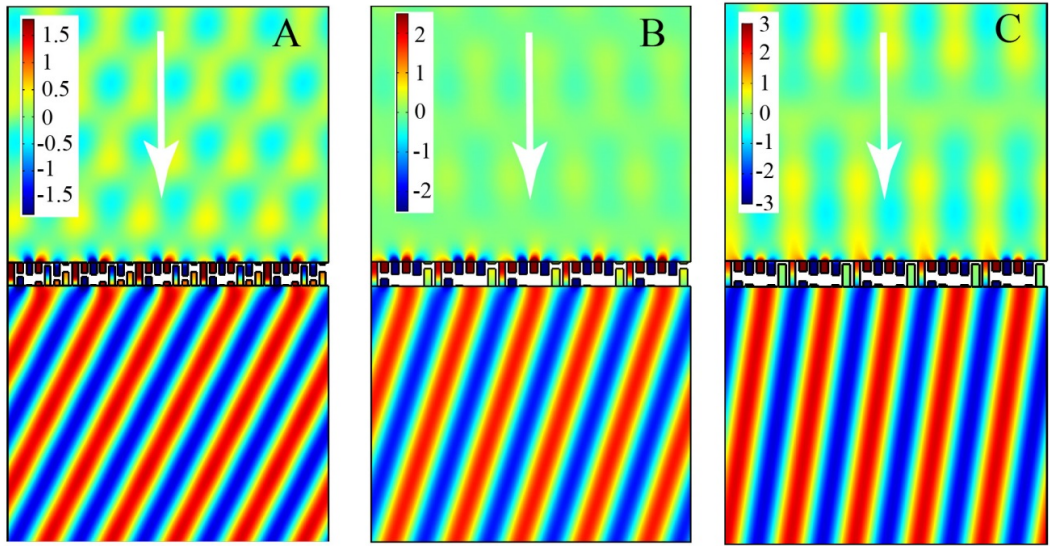

**Fig.2 Field distribution of the scattering wave.** Real part of the pressure field distribution (within the area $0<x<a$ and $y<4a$) of the refractive and reflective wave from the structure. The normally incident plane wave is marked only by the white arrows, the field of them are not shown for clear eyesight. **(A)**, **(B)** and **(C)** are for the structures with refractive angles as $\theta_t=63°$, 72° and 81°, respectively.

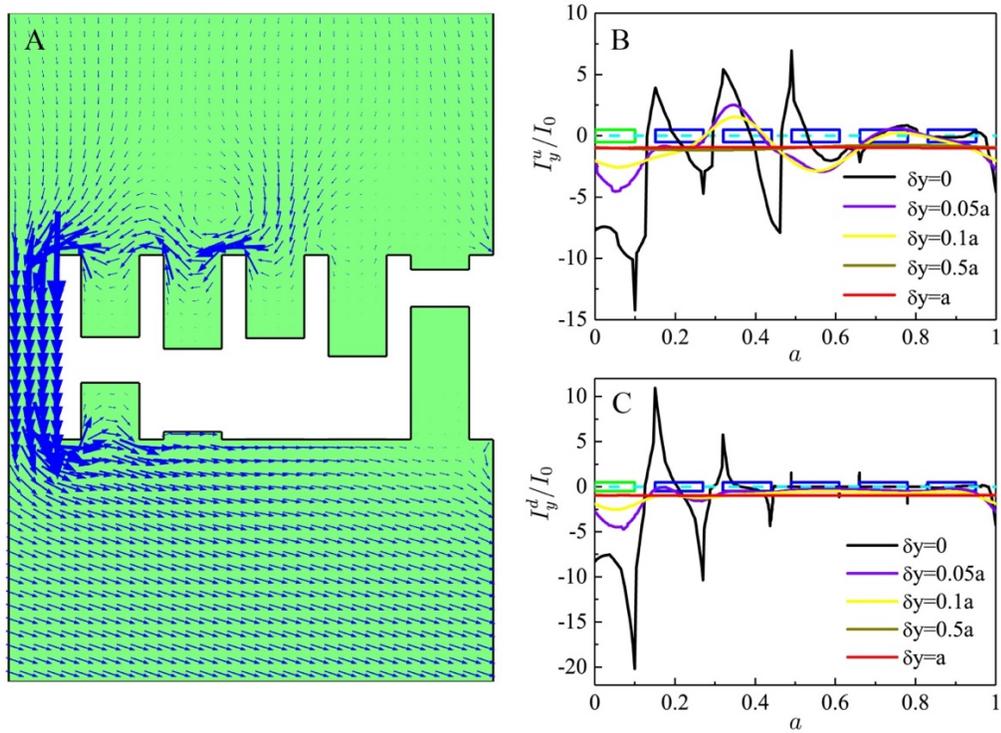

**Fig. 3. Local intensity of the field.** **(A)** Local intensity vector distribution for the total field of the structure with $\theta_t=72°$. The incident plane wave is from $-y$ direction. **(B)** and **(C)** gives the power flow along the $+y$ and $-y$ directions at different distances ($\delta_y$) away from the upper or lower interfaces, respectively.

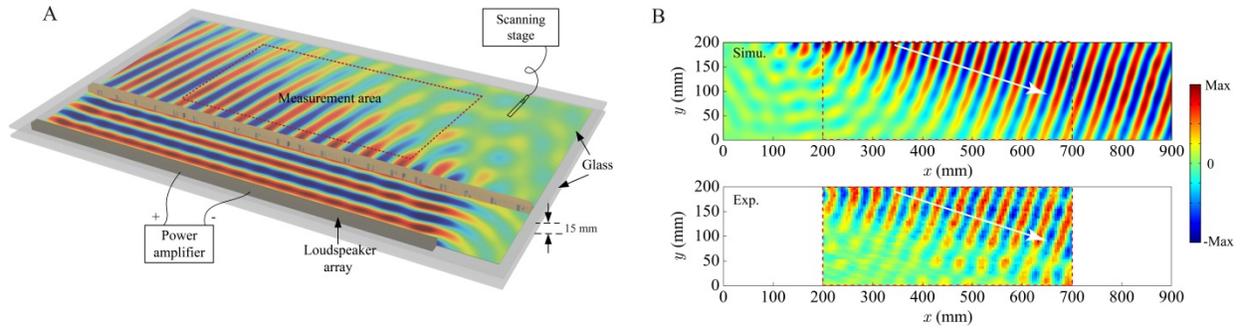

**Fig. 4 Experimental setup and the measured scattering field. (A)** Schematic representation of the experimental setup. **(B)** Pressure wave distribution of the refractive wave from the structure for $\theta_t=72°$. The structures contains totally 20 periods. The upper panel is the result simulated by finite element method, and the lower panel is the measured data in the areas marked by the red box in the corresponding upper panel. The white arrows show the directions of the transmitted wave.